# Performance-Optimum Superscalar Architecture for Embedded Applications

Mehdi Alipour[1], Mostafa E. Salehi[2]*

[2]Assistant Professor, Electrical, Computer, and IT Dept., Islamic Azad University, Qazvin Branch, Qazvin 34185-1416 Iran.
[1]Graduate Student, Electrical, Computer, and IT Dept., Islamic Azad University, Qazvin Branch, Qazvin 34185-1416 Iran.
*Emaild id: m.e.salehi@qiau.ac.ir
[2]Emaild id: mehdi.alipour@qiau.ac.ir

**ABSTRACT**

*Embedded applications are widely used in portable devices such as wireless phones, personal digital assistants, laptops, etc. High throughput and real time requirements are especially important in such data-intensive tasks. Therefore, architectures that provide the required performance are the most desirable. On the other hand, processor performance is severely related to the average memory access delay, number of processor registers and also size of the instruction window and superscalar parameters. Therefore, cache, register file and superscalar parameters are the major architectural concerns in designing a superscalar architecture for embedded processors. Although increasing cache and register file size leads to performance improvements in high performance embedded processors, the increased area, power consumption and memory delay are the overheads of these techniques. This paper explores the effect of cache, register file and superscalar parameters on the processor performance to specify the optimum size of these parameters for embedded applications. Experimental results show that although having bigger size of these parameters is one of the performance improvement approaches in embedded processors, however, by increasing the size of some parameters over a threshold value, performance improvement is saturated and especially in cache size, increments over this threshold value decrease the performance.*

## 1. INTRODUCTION

Embedded systems are designed to perform customised functions often with real-time constraints, while general-purpose processors are designed to be flexible and meet a wide range of end-user requirements. Embedded systems are used to control many devices in common use today [1], such that more than 10 billion embedded processor have been sold in 2008 and more than 10.75 billion in 2009 [2]. Since embedded applications have become more complex, embedded processors have motivated a wide area in research and encourage the designers to optimise theses processors. In addition, embedded processor designers have strict power and area limitations. Recently multiprocessors, very long instruction word (VLIW) and multi-issue superscalar processors satisfy the high performance requirements. However, these architectures have complex design flows [3-7]. Multiprocessors are most often used in recent researches [4-6, 8-11]. Multiprocessors have many computational resources on chip and in comparison with high frequency single processors, can reach higher performance in lower power consumption by running concurrent threads and cores in lower frequencies [7, 11-13, 16]. Based on the researches on embedded applications and processors, multiprocessors that have multithreaded architectures can deliver better performance, exploiting both instruction and thread level parallelism [13-16], this is the main reason why recent researches on embedded processors are base on multiprocessors that run multiple threads [10,13-19].

Overall, the performance of a multi-thread processor is related to many parameters. One of the most important parameter is the sharing strategy applied to the resources shared between threads [20]. Dynamic sharing method has been used in [14, 21, 22] in which threads participate in a competition for accessing the recourses. Static approaches have also been used by designers in which each thread has dedicated recourses and in comparison, static approaches are better than dynamic approaches when the number of threads is low and chip resources do not have a limited budget [6, 8, 11]. An appropriate approach to implement a multi-thread chip is re-implementing a single thread processor with multiple threads. However, there are many limitations in designing high performance single thread processor for embedded applications and there is no standard to convert a single-thread to a multi-thread processor. So, design space exploration is required to find the optimum design parameter values [32-36]. However, in these researches, there is no limitation on the upper bound size of design parameters. Since performance per area and per power are of the most important parameters in designing an embedded processor, in this paper, we explore the design space of a single thread processor to find the optimum size of architectural parameters such as cache, register file, superscalar parameters -instruction window, reorder buffer (ROB) size, instruction queue (IQ) and load store queue (LSQ).

## 2. RELATED WORK

Recent researches are based on comparisons to single-thread single-core processors. In other words, multi-thread processors are the heir to single-thread processors. So exploring important parameters like cache, register file, ROB, IQ, LSQ and branch prediction is required and is the purpose of this paper. Generally, one approach for improving the performance of general purpose and embedded processors is increasing the cache size [13, 15, 17, 26, 27]. However, larger caches consume more power and area. Therefore, it is necessary to find the specific size of the cache and other parameters that create tradeoffs between performance, area and power consumption in embedded designs. On the other hand by considering performance per area parameter that is one of the most important parameter in embedded processors, performance improvements with minimum area overhead is one of the most important needs of a high performance embedded processor.

Recently multi-thread processors are used for designing fast embedded processor [11, 12, 16, 19]. In [23] fine-grain multi-threading based on Markov model has been applied. Analytical Markov model is faster than simulation and has dispensable inaccuracy. In this model states are inactive threads and transition between them are caused by cache misses. In [10], miss rate is reported as the major reason of performance degradation of thread level parallelism (TLP). In [6] thread criticality prediction has been used and for better performance, resources are granted to thread that have higher L2 cache miss rates. These threads are called the most critical threads. To improve the performance of embedded processors in packet-processing applications, in [9, 24, 25] direct cache access (DCA) has been applied as an effective approach. In [16], simultaneous multithread (SMT) processors are introduced as the base processors for network processors and cache miss rates are considered for evaluating the performance improvements. Using victim caches is also mentioned as an approach for performance improvement in multi-thread processors [11]. In multi-thread processors, making the cache bigger increases the cache access delay, and with large caches, executing the memory operations causes pipeline blocking and therefore performance degradations. Hence, the tradeoff between cache size and number of threads is an important design concern. In this paper by considering the fact that larger caches have longer access delays [28], optimum size of the cache is explored for embedded applications. Another important concern in the design of embedded processors is register file size. Similar to cache, size of this module has fundamental effects on the processor performance. To improve the performance of embedded processors, large register files should be used. However, large register file occupies large area and increases the critical path [29, 30] therefore, obtaining the optimum size of the register file is the second purpose of this paper.

Generally, high performance processors are implemented with multi-issue architectures and out of order (OOO) instruction execution [32, 38-41]. On the other hand, since register files are shared in multi-thread processors, larger register files leads to better performance [30, 31, 42]. The effect of register file size on the performance of SMT processors has been studied in [33] and large register files have been proposed. In multi-thread superscalar processors instructions are executed out of the programme order. Therefore, superscalar parameters have an effective impress on the execution of instructions. Superscalar parameters such as ROB, LSQ, IQ, type of branch prediction, and register renaming contribute in the qualification of out of order (OOO) execution [20, 34, 35, 36]. In dynamic scheduled micro-architectures, the execution window size is determined primarily by the capacity of the processor's instruction queue (IQ), which holds decoded instructions until their operands and appropriate function units are available [37]. ROB is used to return the instructions back to the programme order before OOO execution [38]. In OOO processors LSQ is designed to reduce the consecutive accesses to cache and preserve the order of memory operation by storing the instructions competing for memory considering the programme order. By increasing the size of any of these parameters, limited performance improvement can be met, but it does not always work. For example, because of the size, access rate, operation form and associatively implementation, IQ is one of the high power consuming parts, so, any change in the size of IQ, has direct effect on total power consumption [35].

By increasing both the clock rate of a processor and the number of pipeline stages, the micro-architecture parameters also have to tolerate changes in their size [36]. An important question is that what is the optimum and best size for these parameters? This question has created a research area on exploring the size of superscalar parameters [20, 37, 39, 40]. In this field, studies on LSQ are less than the others. Whereas, another effect of higher clock frequency is long delays for accessing memory, which introduces a gap between the performance of processor and memory, and consequently increases the number of instructions competing for memory and hence, increases the pressure on LSQ. This is the reason for choosing large sizes for LSQ. Indeed the approaches to reduce the size of cache, register file, IQ, LSQ and ROB without considering the nature of programmes, generally make higher number of rival instructions on the memory in applications with memory bottleneck [30, 31, 34, 37-47]. So, in this paper by considering this effect, we explore the architectural level parameters of embedded applications to present a performance efficient embedded processor.

## 3. BENCHMARK

The aim of this paper is to calculate the optimum architectural parameters. We have applied our design space exploration (DSE) on heterogeneous applications from PacketBench [48] and MiBench [49]. PacketBench is a good platform to evaluate the workload characteristics of network processors. It reads/ writes packets from/ to real packet traces, and manages packet memory, and implements a simple application programming interface API. This involves reading and writing trace files and placing packets into the internal memory data structures used by PacketBench. PacketBench applications are categorised in three parts: *(i)- IP forwarding* which is corresponding to current internet standards. *(ii)- Packet classification* which is commonly used in firewalls and monitoring systems. *(iii)- Encryption,* which is a function that actually modifies the entire payload of the packet. Specific applications that we used from each category are IPv4-Lctrie, Flow-Classification and IPSec respectively. IPv4-trie performs RFC1812-based packet forwarding. This implementation is derived from an implementation for the Intel IXP1200 network processor. This application uses a multi-bit Trie data structure to store the routing table, which is more efficient in terms of storage space and lookup complexity [48]. Flow classification is a common part of various applications such as firewalling, NAT and network monitoring. The packets passing through the network processor are classified into flows which are defined by a 5-tuple consisting of IP source and destination addresses, source and destination port numbers and transport protocol identifier. The 5-tuple is used to compute a hash index into a hash data structure that uses link lists to resolve collisions [48]. IPSec is an implementation of the IP Security Protocol [49], where the packet payload is encrypted using the Rijndael Advanced Encryption Standard (AES) algorithm [50]. This is the only application where the packet payload is read and modified.

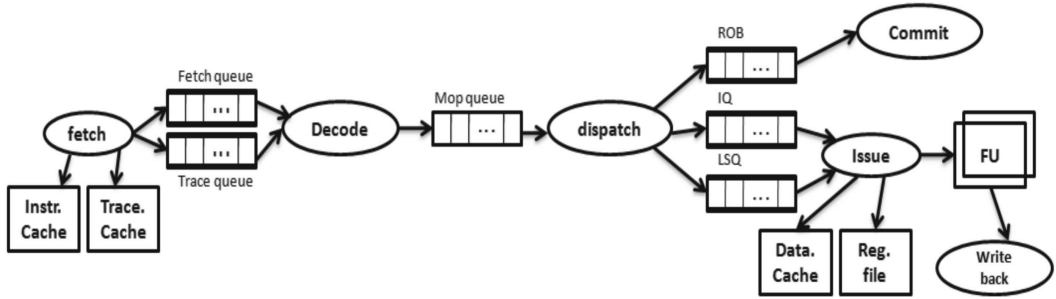

**Figure 1:** *Processor pipeline of multi2sim simulator [46]*

**Table 1:** *The most important parameters used in CACTI*

|  | L1 cache | L2 cache |
|---|---|---|
| Cache size, Cache line size, Associatively | Variable | Variable |
| Number of banks | 1 | 1 |
| Technology node (nm) | 90nm | 90nm |
| Read/write ports | 1 | 1 |
| Exclusive read ports | 0 | 0 |
| Exclusive write ports | 0 | 0 |
| Change tag | No | No |
| Type of cache | Fast | normal/serial |
| Temperature (K) | 300-400 | 300-400 |
| RAM cell/transistor type in data array | ITRS-HP | Global |
| RAM cell/transistor type in tag array | ITRS-HP | Global |

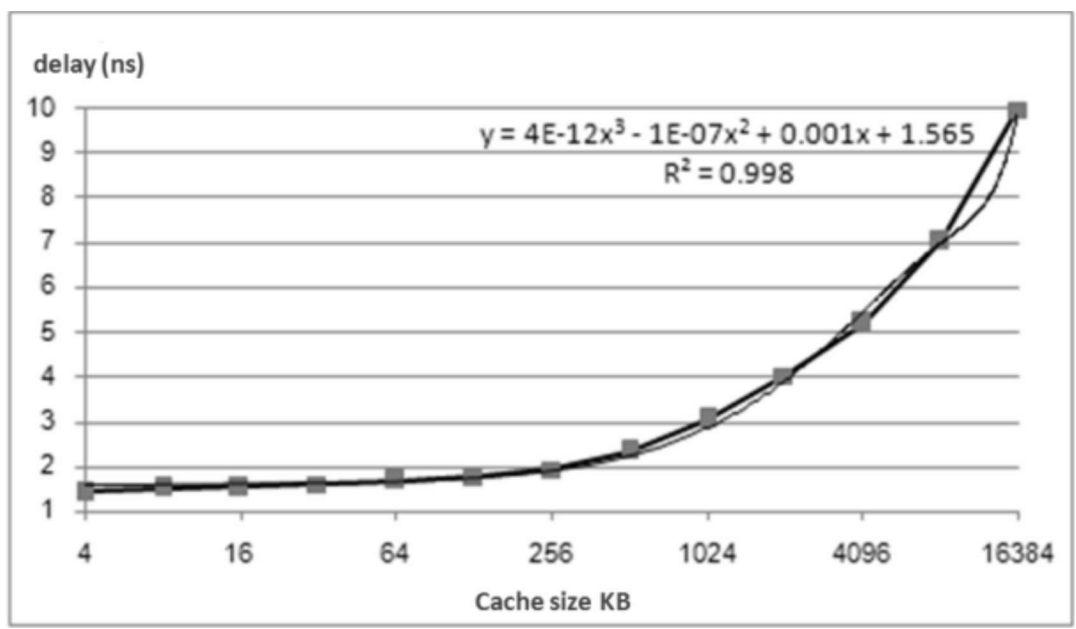

**Figure 2:** *Effect of cache size on cache access delay*

MiBench is a combination of six different categories. We have selected three of them: *(i)- Dijkstra* from network category, *(ii)- Susan (corners)* from automotive and industrial control category, and *(iii)- String-search* from office category. The *Dijkstra* benchmark constructs a large graph in an adjacency matrix representation and then calculates the shortest path between each pair of nodes using repeated executions of *Dijkstra's* algorithm [49]. *Susan* is an image recognition package. It was developed for recognising corners and edges in magnetic resonance images of the brain [49]. *String-search* searches for given words in phrases using a case insensitive comparison algorithm.

## 4. SIMULATION METHODS AND RESULTS

The purpose of this paper is to evaluate optimum size of cache, register file and superscalar parameters. At first, we describe the methodology to extract proper size of cache. For this purpose, it is necessary to configure the simulator in the way that just the size of cache is the parameter that affects the performance. So, for each application the execution number of the main function is calculated in different sizes of L1 and L2 caches.

For this purpose we made changes in some parts of the simulator source code to calculate the cycles that are used to execute the main function of each application. To calculate the start address and end address of the main function, we have disassembled the executable code of each benchmark application and extract these addresses and then these parameters are back annotated to *commit.c* and *processor.h* file of Multi2sim simulator running a thread of the selected application. By these changes we can calculate the number of x86 instructions and macroinstructions and also count number of the cycles for specific function. The processor pipeline of Multi2sim simulator is also shown in Figure 1. The second step is to run the simulator with different cache sizes. However, the worthwhile point is that although based on the recent researches that recommend doubling the cache size for improving the performance of a processor, during doubling the cache size, important parameters like area power and cache access delay must be considered. For this purpose we have used CACTI 5.0 [28], a tool from HP that is a platform to extract parameters relevant to cache size considering fabrication technology. Most important parameters that we used in this research are listed in Table 1.To compare the performance based on the cache size, extracted results from CACTI (L1 and L2 cache access delay) are back annotated to Multi2sim. In this way when the cache size is changed, actual cache access delays are considered.

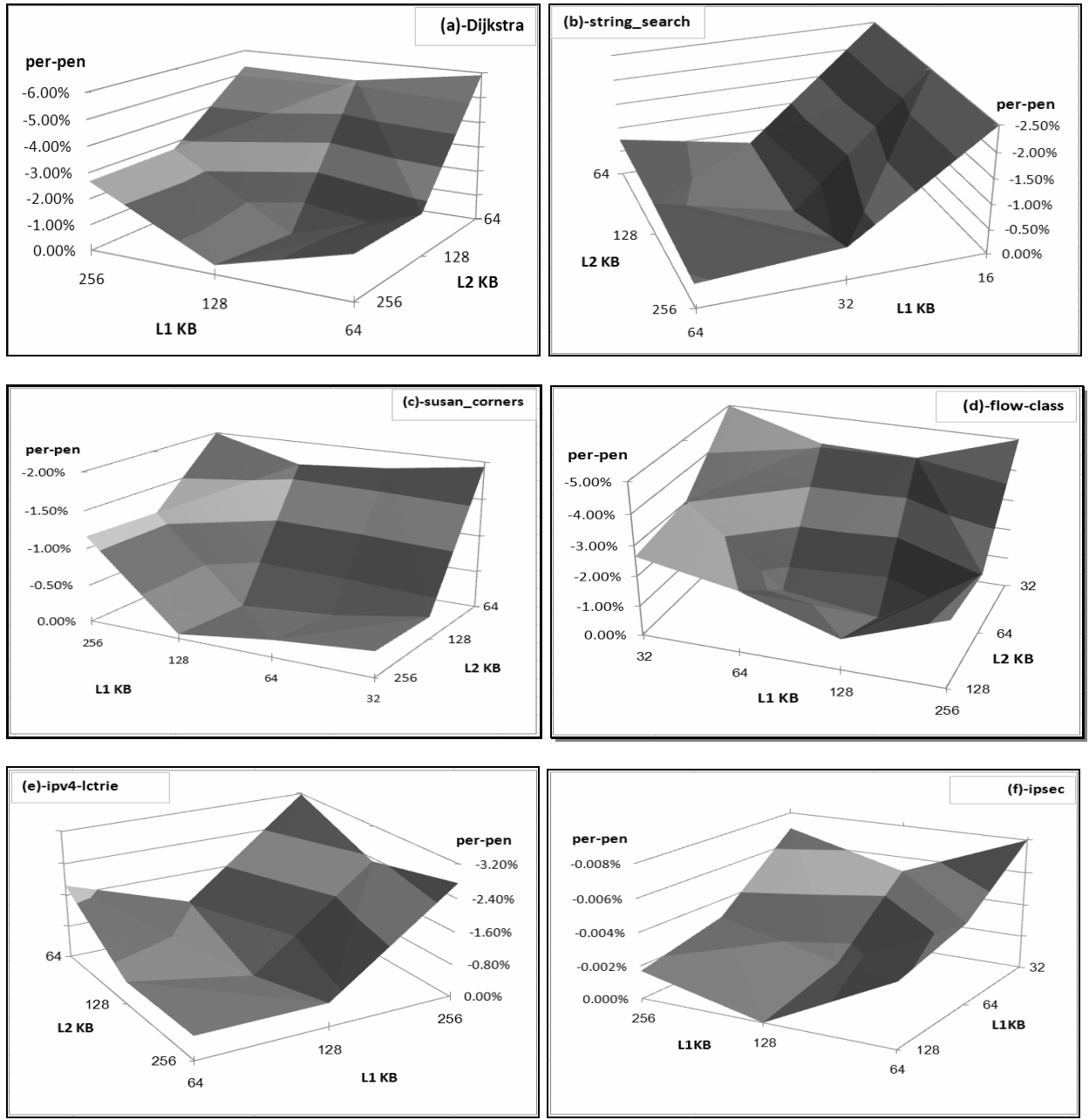

**Figure 3:** *Effect of cache size on the performance of embedded applications (a) Dijkstra, (b) String_search (c)Susan.corners (d)flow_class (e)ipv4_lctrie, (f) ipsec*

As can be seen in Figure 2, increasing the cache size, leads to more cache access delays. For exploring the cache size, since here we want to find the best cache size for a single-thread single-core processor for embedded applications, the rest of the simulator parameters are set to the default values, i.e. width of the pipeline stages is one (issue-width =1). Figure 3 shows the extracted results from our simulations.

In this figure the vertical axis (per-pen) shows the performance penalty of related cache size configuration compared to a default size (L1-L2 = 64-128) that is an applicable cache size for embedded processors based on our exploration. Based on these results, by increasing the cache size we can achieve more hit rates, however, because of the longer cache access time for larger caches, from a specific point -which we call the *best cache size* (BCS) in the rest of the paper- performance improvement is saturated and then even decreased. In other words, doubling the cache size always cannot improve the performance. From another point of view, area budget is limited and always we cannot have a large cache, so, by considering the sizes which are smaller and near the *BCS*, performance degradations are negligible (3% in average).

To calculate the optimum size of register file, we have applied the parameters used for calculating the *BCS*. However, to find out just the effect of register file size on the performance, we used the BCS (for both L1 and L2) concluded in the previous section for cache size and run the simulator accordingly. Figure 4 shows the results of register file explorations. In this figure the horizontal axis shows the explored sizes of register file. Each benchmark is represented in a column and right most columns in each size are the average performance penalty of all benchmarks. Value of per_pen in this figure is relative to the default size (# of register =80). Figure 4 shows that although for all applications the best size of register file is almost 72 and above in average, however, in size=48 that is near the half of the best size, performance penalty is lower that -2%. Also this figure shows that reducing the register file size always decreases the performance but sometimes, by doubling the register file size we do not have noticeable performance improvement. So the first point that the highest performance is met will be introduced as the best size for register file. It is worthwhile to say that in Figure 4 the concurrent effect of cache size and register file size can be seen.

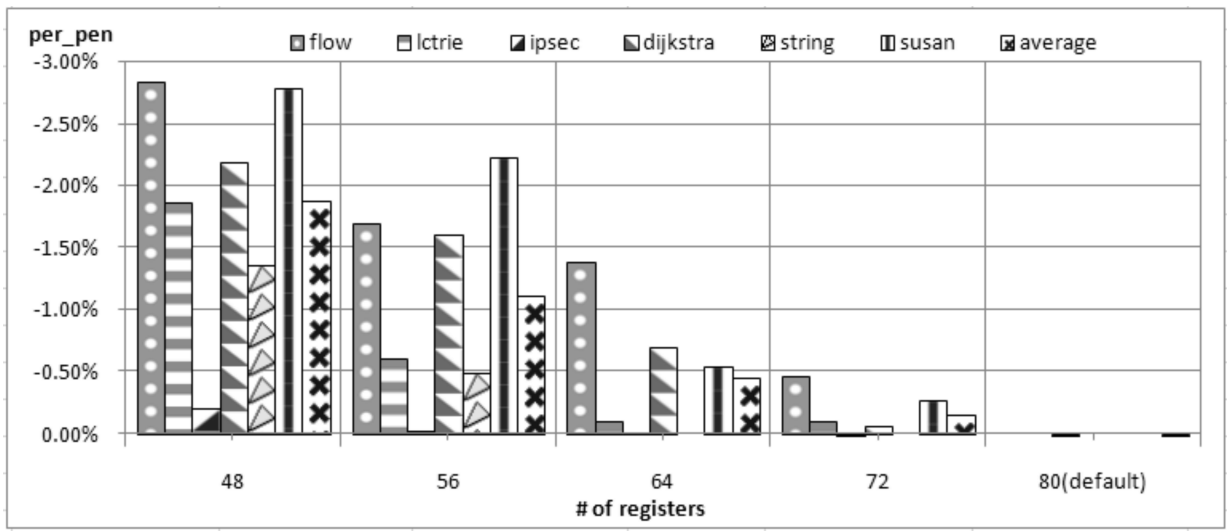

**Figure 4:** *Effect of register file sizes on performance of embedded applications*

From another point of view based on recent researches, multi-thread architectures need more area budget [16, 19, 39]. Furthermore, to meet the best performance, multi-issue architectures require renaming logic, ROB, LSQ, IQ and other OOO components which occupy large area budgets. Based on our simulations, we introduce two points for cache and register file sizes. *(i)-The Best size* that is the first point which has no performance penalty and occupy large area budget and *(ii)-The Optimum size* that has about 2% performance penalty and occupy smaller area budget.

Based on these results, we can deduce that in the optimum sizes of cache (32 to 64KB for L1 and 64 to 128KB for L2) and optimum size of register file (register file size = 56) we can save the area budget of a single thread and instead use more threads for a multi-thread processor. In other words, using the optimum size leads to better performance per area. Using the optimum size of cache and register file, we can make room for extending superscalar parameters (ROB, LSQ and IQ) and hence improve the performance. Figure 5 shows the effects of each superscalar parameter size on the processor performance. Related performance of each size is evaluated according to the default size which is the right most size in the figure. Each benchmark has a dedicated column in each size that shows the *performance penalty* (per_pen) of the application in the specified size. The results show that similar to cache and register file, by doubling the size the performance is not always improved. So from performance point of view, in average, in sizes near half of the best size (the default sizes in this figure), performance penalty for all benchmarks is negligible and considering area and power consumption, lower power and area will be consumed by these parts. Result of Figure 5 (a) show the effects of ROB on performance and indicate that although selected benchmarks are from different categories, they have the same behaviour against ROB variations. It means that a specific size of ROB is applicable for selected embedded benchmarks. Results shows that for these benchmarks, the best size for ROB is 64 with 0.0% performance penalty and the optimum size is 32 or 34 with -1.91% and -1.63% performance penalties in average, respectively. Result of IQ exploration in Figure 5 (b) show the effects of IQ on the performance of different benchmarks and indicate that embedded applications have also the same behaviour in different IQ sizes, so one size can be used for all of these embedded benchmarks. Bigger instruction queue and longer instruction fetch queue can improve the performance of network, industrial and office applications and in the lower IQ sizes network applications gain more performance penalty relatively. So the best point and optimum point for IQ size are 20 (-0.26% pp) and 8 (-1.08% pp), respectively. Result of Figure 5 (c) show the effect of LSQ on performance, and propose 8 as the optimum and 12 as the best size for LSQ for selected embedded applications.

## 5. CONCLUSION

In this paper we have explored the effect of architecture level parameters (cache, register file, ROB, LSQ and IQ) on the performance of the embedded applications and consequently find the performance optimum superscalar architecture for embedded applications. Experimental results show that although having bigger size for mentioned parameters is one of the performance improvement approaches in embedded processors, however, by increasing the size of these parameters over a threshold level, performance improvement is saturated and especially in cache size increments over this threshold point, degrade the performance. We have introduced two points for all of these parameters, the best size which has no performance penalty and the optimum size that has negligible performance penalty and will have good power and area saving for all benchmarks in average. Experiments show that an optimum architecture can be used for all of the selected embedded applications that with the parameter sizes near half of the best size just have about -0.25 performance penalty in average.

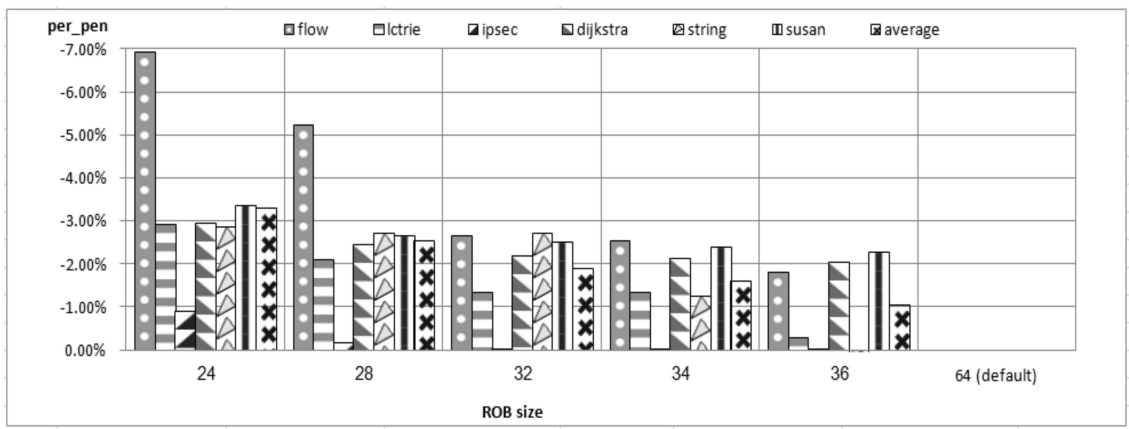

**(a) Performance effects of ROB size**

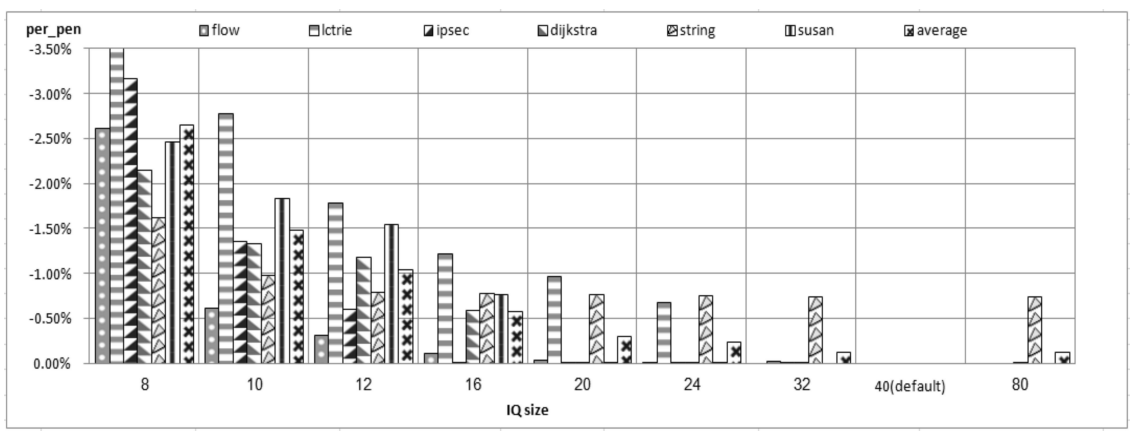

**(b) Performance effects of IQ size**

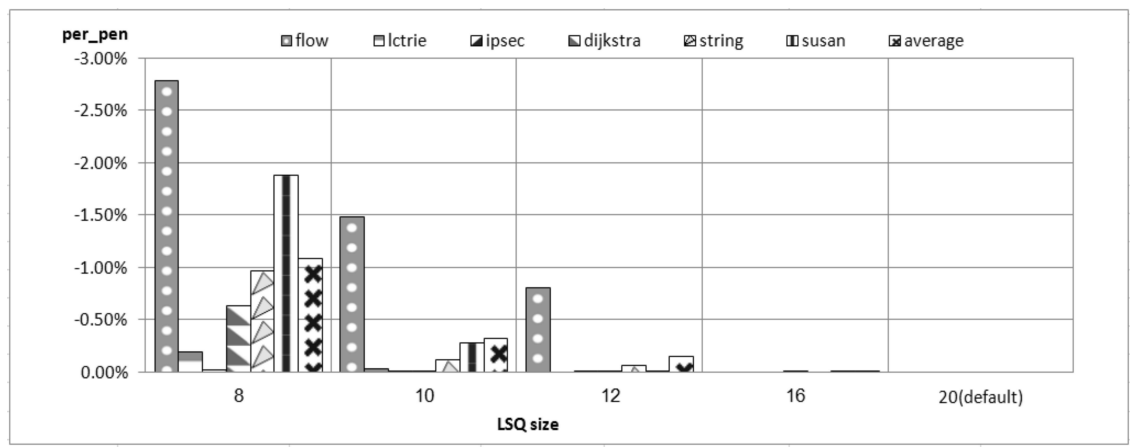

**(c) Performance effects of LSQ size**.

**Figure 5.** *Effect of superscalar parameters on the performance of embedded applications*